\documentstyle[12pt]{article}

\newcommand{\CMP}[1]{{\em Commun. Math. Phys.} {\bf {#1}}}

\newcommand{\NP}[1]{{\em Nucl.Phys.~B} {\bf {#1}}}

\newcommand{\PL}[1]{{\em Phys. Lett.} {\bf {#1}}}
\newcommand{\PR}[2]{{\em Phys. Rev.} {#1} {\bf {#2}}}
\newcommand{\PRL}[1]{{\em Phys. Rev. Lett.} {\bf {#1}}}
\newcommand{\LMP}[1]{{\em Lett. Math. Phys.} {\bf {#1}}}

\newcommand{\Map}{\mbox{{\em Map}}}

\newcommand{\dvp}{\f{d^3 p}{(2\pi)^3}}
\newcommand{\dvx}{d^3 x}
\newcommand{\dvq}{\f{d^3 q}{(2\pi)^3}}
\newcommand{\dvy}{d^3 y}
\newcommand{\intS}{\int_{\R^3}}

\newcommand{\half}{\mbox{$\frac{1}{2}$}}

\newcommand{\U}{{\rm U}}
\newcommand{\su}{{\rm su}}
\newcommand{\gl}{{\rm gl}}
\newcommand{\ul}{\underline}

\newcommand{\eq}{\begin{equation}}
\newcommand{\eqend}{\end{equation}}
\newcommand{\eqa}{\begin{eqnarray}}
\newcommand{\nonueqa}{\begin{eqnarray*}}
\newcommand{\eqaend}{\end{eqnarray}}
\newcommand{\nonueqaend}{\end{eqnarray*}}
\newcommand{\nonu}{\nonumber \\ \nopagebreak}
\newcommand{\bma}[1]{\begin{array}{#1}}
\newcommand{\ema}{\end{array}}
\newcommand{\bc}{\begin{center}}
\newcommand{\ec}{\end{center}}

\newcommand{\Ref}[1]{(\ref{#1})}

\newcommand{\ee}[1]{\mbox{{\rm e}}^{#1}}
\newcommand{\ii}{{\rm i}}
\newcommand{\OO}{{\rm O}}

\newcommand{\vp}{\vec{p}}
\newcommand{\vx}{\vec{x}}
\newcommand{\vq}{\vec{q}}
\newcommand{\vy}{\vec{y}}

\newcommand{\Sym}[1]{\sigma\left({#1}\right)}

\newcommand{\om}{\omega}

\renewcommand{\phi}{\varphi}
\newcommand{\sig}{\sigma}

\newcommand{\Om}{\Omega}

\newcommand{\eps}{\varepsilon}
\newcommand{\Lam}{\Lambda}
\newcommand{\sign}{{\rm sign}}

\newcommand{\R}{{\rm I\kern-.2emR}}
\newcommand{\C}{{\sf C} \! \! \! {\sf I}\:}

\newcommand{\f}{\frac}
\newcommand{\cA}{{\cal A}}

\newcommand{\cL}{{\cal L}}

\newcommand{\ccr}[2]{{[} {#1},{#2} {]} }        
\newcommand{\car}[2]{{\{} {#1},{#2} {\}} }        
\newcommand{\ETC}[2]{{[} {#1},{#2} {]}_{{\rm ETC}}}    

\newcommand{\Tra}[1]{{\rm Tr} \left({#1}\right)}          
\newcommand{\TraL}[1]{{\rm Tr}_\Lam \left({#1}\right)}          
\newcommand{\TraC}[1]{{\rm Tr}_C \left({#1}\right)}          
\newcommand{\tra}[1]{{\rm tr'} \left({#1}\right)}          
\newcommand{\tras}[1]{{\rm tr}_{spin} \left({#1}\right)}          
\newcommand{\trac}[1]{{\rm tr} \left({#1}\right)}          

\newcommand{\pgf}[1]{{\bf {#1}.}}

\newcommand{\Grz}{Gr_2}
\newcommand{\Grp}{Gr_p}
\newcommand{\cuv}{u\leftrightarrow v}
\newcommand{\cXY}{X\leftrightarrow Y}

\newcommand{\g}{\ul{g}}

\newcommand{\gz}{\ul{g}_2}
\newcommand{\gp}{\ul{g}_p}
\newcommand{\dd}{{\rm d}}
\newcommand{\D}{D\!\!\!\!\slash}
\newcommand{\ddd}{\hat{\rm d}}
\newcommand{\slD}{\!\!\!\slash}
\newcounter{saveeqn}
\newcounter{App} 

\newcommand{\alpheqn}{%
\stepcounter{equation}
\setcounter{saveeqn}{\value{equation}}%
\setcounter{equation}{0}%
\renewcommand{\theequation}{\arabic{saveeqn}\alph{equation}} }
\newcommand{\reseteqn}{\setcounter{equation}{\value{saveeqn}}%
\renewcommand{\theequation}{\arabic{equation}} }

\begin{document}
\pagestyle{empty}
\renewcommand{\thefootnote}{\alph{footnote}}
\begin{center}

{\Large \bf (3+1)-Dimensional Schwinger Terms and Non-commutative Geometry}\\
\vspace{1 cm}

{\large Edwin Langmann\footnote{supported in part by "\"Osterreichische
Forschungsgemeinschaft" under contract Nr.\ 09\slash 0019} and Jouko
Mickelsson} \\
\vspace{0.3 cm}
{\em Theoretical Physics, Royal Institute of Technology, S-10044 Sweden}\\
{\em email:} {langmann,jouko@theophys.kth.se}\\
\end{center}

\setcounter{footnote}{0}
\renewcommand{\thefootnote}{\arabic{footnote}}

\begin{abstract}
We discuss 2-cocycles of the Lie algebra $\Map(M^3;\g)$ of smooth,
compactly supported maps on
3-dimensional manifolds $M^3$ with values in a compact, semi-simple Lie
algebra $\g$.  We show by explicit calculation that the
Mickelsson-Faddeev-Shatashvili cocycle $\f{\ii}{24\pi^2}\int\trac{A\ccr{\dd
X}{\dd Y}}$ is cohomologous to the one obtained from the cocycle given by
Mickelsson and Rajeev for an abstract Lie algebra $\gz$ of Hilbert space
operators modeled on a Schatten class in which $\Map(M^3;\g)$ can be
naturally embedded.  This completes a rigorous field theory derivation of
the former cocycle as Schwinger term in the anomalous Gauss' law
commutators in chiral QCD(3+1) in an operator framework.  The calculation
also makes explicit a direct relation of Connes' non-commutative geometry
to (3+1)-dimensional gauge theory and motivates a novel calculus
generalizing integration of $\g$-valued forms on 3-dimensional manifolds to
the non-commutative case.

\end{abstract}

\newpage
\pagestyle{plain}
\setcounter{page}{1}

{\bf 1.  Introduction.} Infinite dimensional Lie algebras $\Map(M^d;\g)$ of
smooth, compactly supported maps from a $d$-dimensional manifold $M^d$ to a
compact, semi-simple Lie algebra $\g$ (e.g.\ $\g=\su(N)$) are closely
related to ($d$+1)-dimensional quantum field theory (QFT).  One strong
motivation for studying projective representations of these algebras is the
hope that they could lead to progress in the understanding of the
non-perturbative structure of associated QFT models.  This has been indeed
so for $d=1$: the by-now well-understood representation theory of the loop
algebras $\Map(S^1;\g)$ has played a crucial role in recent spectacular
progress in (1+1)-dimensional QFT (e.g.  conformal QFT(1+1); for a recent
construction of QCD(1+1) with massless quarks based on the representation
theory of loop algebras see \cite{LS}).

One natural interpretation of $\Map(M^d;\g)$ is as Lie algebra of the group
of static gauge transformations of a Yang-Mills gauge theory on space-time
$M^d\times\R$.  It is then natural to consider also the set $\cA(M^d)$ of
all (static) Yang-Mills field configurations $A$ on $M^d$, i.e.\ $A$ are
the $\g$-valued, compactly supported 1-forms on $M^d$.  For $d=3$ an
extension of $\Map(M^3;\g)$ is given by the Mickelsson-Faddeev-Shatashvili
cocycle \cite{M,FS}
\eq
\label{MFS}
c_{MFS}(X,Y;A)=\f{\ii}{24\pi^2} \int_{M^3} \trac{A\ccr{\dd X}{\dd Y}}
\eqend
($X,Y\in \Map(M^3;\g)$, $A\in \cA(M^3)$; we use the standard notation for
forms on $M^3$ ($\dd$ is the exterior derivative etc.) suppressing the wedge
product, and implicitly assume a representation of $\g$ in some
$\gl(N)$ (algebra of complex $N\times N$ matrices) acting on
$\C^N=\C^N_{color}$ where ${\rm tr}$ is the usual trace of $N\times N$
matrices).

There are ``big'' abstract Lie algebras $\gp$ of operators on a Hilbert
space modeled on Schatten classes which play a central role in the
mathematical investigation of $\Map(M^d;\g)$.  The motivation for
introducing $\gp$ is that it naturally contains $\Map(M^d;\g)$ for {\em
any} `nice'\footnote{$C^\infty$ manifold with a Riemannian- and a spin
structure} $d$-dimensional manifold $M^d$ if $p=(d+1)/2$, and that it is
possible to develop the representation theory of $\gp$ as a whole and
obtain the ones of $\Map(M^d;\g)$ by restriction from that \cite{MR,L}.
Actually, for $p>1$ the representation theory of $\gp$ requires to
introduce another ``big'' set of operators $\Grp$ --- the so-called
Grassmannian --- modeled on the same Schatten class as $\gp$.  From a
physical point of view this is quite natural as one can naturally embed the
sets $\cA(M^d)$ of Yang-Mills configurations in $\Grp$ if $p=(d+1)/2$, and
there is a natural action of $\gp$ on $\Grp$ generalizing the gauge
transformations by which elements of $\Map(M^d;\g)$ act on $\cA(M^d)$.  It
is also interesting to note that these very Lie algebras $\gp$ play a
fundamental role in Connes' non-commutative geometry \cite{Connes}.

To define $\gp$ and $\Grp$ one considers a separable Hilbert space $h$
which is decomposed in a direct sum of two infinite dimensional, orthogonal
subspaces, $h=h_+\oplus h_-$ (we recall that abstractly, all such Hilbert
spaces $h$ are essentially --- up to unitary equivalence --- the same).  Such a
decomposition is uniquely determined by the operator $\eps$ on $h$ which is
$+1$ on $h_+$ and $-1$ on $h_-$, $h_\pm=\f{1}{2}(1\pm \eps)h$.
Then $\gp$ is defined as the Lie algebra of all bounded operators on $h$
such that $(\ccr{\eps}{u}^*\ccr{\eps}{u})^p$ is trace class
($*$ is the Hilbert space adjoint; we recall that an operator $a$ on $h$
is {\em trace class} if $\sum_n |<f_n, af_n>|$ is finite for any complete
orthonormal basis  $\{ f_n \}$ in $h$, and then its {\em Hilbert space
trace} $\Tra{a}\equiv \sum_n <f_n, a f_n>$ exists, i.e.\ it is finite and
independent of  $\{ f_n \}$ \cite{RS1}).

To explain the embeddings of $\Map(M^d;\g)$ in $\gp$ and $\cA(M^d)$ in
$\Grp$ we consider chiral fermions on space-time $M^d\times\R$ coupled to
an external Yang-Mills field $A\in\cA(M^d)$. Then the Gauss' law
generators $G(X)$ implementing the infinitesimal gauge transformations
$X\in\Map(M^d;\g)$ in the physical Hilbert space of the fermions should
obey equal-time commutators of the following form,
\eq
\label{ETC}
\ETC{G(X)}{G(Y)}=G(\ccr{X}{Y}) + S_{d+1}(X,Y;A)
\eqend
with a Schwinger terms $S_{d+1}$ satisfying a 2-cocycle relation due to
the Jacobi identity for the equal-time commutator \cite{FS}.
For $d=3$ cohomological arguments suggest that this Schwinger term
should be equal (up to a boundary) to the MFS cocycle \Ref{MFS} \cite{FS}.

To explicitly construct these Gauss' law generators, one can start with
the Hilbert space $h$ of 1-particle states of the chiral fermions, i.e.\
$h=L^2(M^d)\otimes V_{spin}\otimes\C^N_{color}$ with $V_{spin}$ a vector
space carrying the spin structure. Then the 1-particle time evolution of
the fermions is determined by the Weyl operator $\D_A$ in the external
field $A$. This is a self-adjoint operator on $h$, and it provides a
natural splitting of $h$ in positive- and negative energy states,
$h=h_+^A\otimes h_-^A$ with $\D_A\geq 0$ ($<0$) on $h_+^A$ ($h_-^A$). It
is this splitting which determines the physical Hilbert space of the
fermions (one has to fill up the Dirac sea corresponding to the negative
energy states). Then $F_A$ is defined to be the operator which is $\pm 1$
on $h_\pm^A$, and $\eps=F_0$ (no external field). Note that the mapping
$A\mapsto F_A$ is continuous along gauge orbits (it has discontinuities
only for those configurations $A$ where an eigenvalue of $\D_A$ crosses
zero).

Infinitesimal gauge transformations $X\in\Map(M^d;\g)$ naturally correspond
to self-adjoint operators on $h$, $(Xf)(\vx)=X(\vx)f(\vx)$ for all $f\in
h$ (to simplify notation we use the same symbol for $X\in\Map(M^d;\g)$
and the corresponding operator on $h$). The basic result implying the
embedding referred to above is that $F_A\in\Grp$ and $X\in\gp$ for all
$A\in\cA(M^d)$ and $X\in\Map(M^d;\g)$ if $p\geq (d+1)/2$, see e.g.\ \cite{MR}.

{}From an abstract point of view, every $u\in\gp$ corresponds to an
infinitesimal fermion transformation and every $F\in\Grp$ to a fermion
Dirac sea, and it is natural to consider implementors $G(u)$ for {\em
all} $u\in\gp$ and $F\in\Grp$ satisfying
\eq
\label{GLabs}
\ETC{G(u)}{G(v)}=G(\ccr{u}{v}) + c_p(u,v;F)
\eqend
where $c_p$ is a 2-cocycle. Indeed, the very definitions of $\gp$ and
$\Grp$ characterize a certain degree of divergence and thus determine a
regularization procedure adequate for this type of divergence \cite{L,M3}.
Moreover, this regularization procedure is uniquely determined by the
2-cocycle $c_p$ up to a coboundary $\delta b$,
\alpheqn
\eq
\label{b1}
(\delta b)(u,v;F)\equiv b(\ccr{u}{v};F) -\cL_u b(v;F) + \cL_v b(u;F)
\eqend
with the Lie derivative $\cL_u$ acting on functions $f(F)$ as
\eq
\cL_u f(F)\equiv \left.\f{1}{\ii}\f{\partial}{\partial t} f(\ee{-\ii u
t}F\ee{\ii ut})\right|_{t=0} .
\eqend
\reseteqn
The change $c_p\to c_p+ \delta b$ corresponds to a {\em finite} (i.e.\
trivial) change of the regularization.  It is worth pointing out that this
abstract construction is not only convenient mathematically but also
natural from the physical point of view: Besides the infinitesimal gauge
transformation, $\gp$ contains also other operators of interest for
$(d+1)$-dimensional gauge theories with fermions (see e.g.\ \cite{L}), and
the mathematical construction of the algebra \Ref{GLabs} should therefore
provide a general procedure adequate for (ultra-violet) divergences in
the matter sector of such theories.

For $p=2$ (corresponding to $d=3$) the natural extension of $\gz$
is given by the Mickelsson-Rajeev cocycle \cite{MR}
\eq
\label{MR}
c_{MR}(u,v;F) =
-\f{1}{8}\TraC{(F-\eps)\ccr{\ccr{\eps}{u}}{\ccr{\eps}{v}} }
\eqend
($u,v\in \gz$, $F\in \Grz$) where we introduced the
conditional trace
\eq
\TraC{a}\equiv \f{1}{2}\Tra{a+ \eps a\eps}
\eqend
which exists and is finite for all operator $a$ on $h$ so that $a+\eps
a\eps$ is trace class . We call such operators {\em conditionally trace
class}.  Note that $\TraC{a}=\Tra{a}$ for all trace class operators $a$.
(We note that the operator $(F-\eps)\ccr{\ccr{\eps}{u}}{\ccr{\eps}{v}}$ for
$F\in\Grz$, $u,v\in\gz$ is {\em not} trace class but only conditionally
trace class in general \cite{L}, hence only its conditional trace exists.
The necessity to use ${\rm Tr}_C$ and not ${\rm Tr}$ in the formula for the
MR cocycle has not been made sufficiently clear in \cite{MR,L}.)

In \cite{MR} a cohomological argument was given that the MFS-cocycle
should be equivalent to the MR-cocycle, i.e.\ given the natural
emdeddings of $\Map(M^3;\g)$ in $\gz$, and $A\mapsto
F_A$ of $\cA(M^d)$ in $\Grp$,
\eq
c_{MR}(X,Y;F_A) = c_{MFS}(X,Y;A) + (\delta b)(X,Y;A)
\eqend
for some boundary $\delta b$.

In this paper we prove by explicit calculation that this is true.
To avoid technicalities we restrict ourselves to the simplest case
$M^3=\R^3$ (we recall that all mappings $X$, $Y$, $A$ considered
have compact support).  The extension of our result to arbitrary manifolds
$M^3$ can then be done using basic results on symbol calculus on manifolds
\cite{symbol}.

We believe that this calculation is interesting for two reasons.  Firstly,
in combination with the results in \cite{L} it provides a {\em rigorous}
derivation of the MFS-cocycle in the anomalous commutators of the Gauss'
law generators in chiral QCD(3+1) in an operator framework, the Yang-Mills
field being treated as external, non-quantized field.  (A different
solution to this problem was recently explained in \cite{M1}.) Though
several field theory derivations of this result exist in the literature
(using the BJL-limit to define equal time commutators, e.g.\
\cite{BJL}, or Berry's phase, e.g.\ \cite{Berry}, none of these is very
satisfactory
from a more mathematical point of view\footnote{An earlier indication of
non-vanishing Schwinger terms was obtained in a perturbative computation of
vacuum expectation values of hadronic currents in external $\U(1)$ gauge
field \protect{\cite{JJ}}}.  Secondly (as we discuss in more detail in the
final paragraph), the calculation shows very explicitly a natural relation
of non-commutative geometry (NCG) \cite{Connes} to (3+1)-dimensional
Yang-Mills gauge theory and motivates a new generalization of the
integration calculus of forms to the non-commutative case.  It has been
repeatedly pointed out by Connes that NCG should provide an appropriate
mathematical framework for formulating and studying {\em quantum} gauge
theory without perturbation theory.  To our knowledge this program has not
yet lead to many new results (one result in this direction is Rajeev's
universal Yang-Mills theory \cite{Rajeev}).  We therefore believe that the
study of Lie algebras $\Map(M^d;\g)$ by extending to operator algebras
$\gz$ provides a very interesting example where the NCG point of view is
successfully used for getting deeper insight in QFT divergences arising in
a gauge theory.  The present paper can be regarded as an attempt to bridge
the gap between this abstract, mathematical approach and more standard
particle physics methods for the physically relevant case $d=3$.

{\bf 2.  Calculation.} \pgf{a} Our Hilbert space is
$h=L^2(\R^3)\otimes\C^2_{spin}\otimes\C^N_{color}$, and the free Weyl
operator can be represented as \footnote{repeated
indices $i\in\{1,2,3\}$ are sumed over throughout}
 $\D_0=(-\ii)\partial_i\sig_i$ where $\sig_i$
are the Pauli spin matrices acting on $\C^2_{spin}$ and
$\partial_i=\partial/\partial x^i$.  For our calculation we need some basic
facts about symbol calculus \cite{symbol}.  We recall that every
pseudodifferential operator (PDO\footnote{all operators of interest to us are
PDOs}) $a$ on $h$ can be represented by its {\em symbol} $\sig(a)(\vp,\vx)$
which is a $\gl(2)_{spin}\otimes\gl(N)_{color}$-valued function on phase
space $\R^3\times\R^3$ and defined such that for any $f\in h$,
\eq
\label{symb}
(af)(\vx) = \int\dvp\,
\ee{-\ii\vp\vx}\sig(a)(\vp,\vx)\hat f(\vp)
\eqend
where $\hat f(\vp) = \int\dvp\, \ee{\ii \vp\vx}f(\vx)$
denotes the Fourier transform of $f$. It follows then that
\eq
\label{product}
\sig(ab)(\vp,\vx)= \intS\dvq\intS\dvy\ee{\ii(\vx-\vy)(\vp-\vq)}
\sig(a)(\vq,\vx)\sig(b)(\vp,\vy),
\eqend
and for $a$ trace-class,
\eq
\Tra{a} =
\intS\dvp\intS\dvx\, \tra{\sig(a)(\vp,\vx)}
\eqend
where ${\rm tr'}={\rm tr}_{spin}{\rm tr}_{color}$.  Especially,
$\sig(\eps)(\vp,\vx)= \f{p\slD}{p}\equiv \eps(\vp)$ where $p\slD\equiv
p_i\sig_i$ and $p\equiv |\vp|$, and $\sig(X)(\vp,\vx)=X(\vx)$ for all
$X\in\Map(\R^3;\g)$.

All operators $a$ of interest to us allow an asymptotic expansion
$\sig(a)\sim\sum_{j=0}^{\infty}\sig_{-j}(a)$ where $\sig_{-j}(a)(\vp,\vx)$ is
homogeneous of degree $-j$ in
$\vp$,\footnote{i.e.\ $\sig_{-j}(a)(s\vp,\vx)=s^{-j}\sig_{-j}(a)(\vp,\vx)$ for
all $s>0$} and it goes to zero like $p^{-j}$ for $p\to\infty$.
We write
\eq
\sig(a)(\vp,\vx)=\sum_{j=0}^{n}\sig_{-j}(a)(\vp,\vx) + \OO(p^{-n-1})
\eqend
for all integers $n$.  Moreover, eq.\ \Ref{product} has an asymptotic
expansion in powers of $p^{-1}$,
\eq
\label{as}
\sig(ab)(\vp,\vx) \sim \sum_{n=0}^{\infty} \f{(-\ii)^n}{n!}
\f{\partial^n\sig(a)(\vp,\vx)}{\partial p_{i_1}\cdots\partial p_{i_n}}
\f{\partial^n\sig(b)(\vp,\vx)}{\partial x_{i_1}\cdots\partial x_{i_n}}.
\eqend
This allows to determine the asymptotic expansion of $\sig(ab)$ from the ones
of $\sig(a)$ and $\sig(b)$.  Especially if $\sig(a)$ is $\OO(p^{-n})$ and
$\sig(b)$ $\OO(p^{-m})$ then $\sig(ab)$ is $\OO(p^{-(n+m)})$.

In our calculation we shall only need the leading terms of the asymptotic
expansion of the symbols of $\ccr{\eps}{X}$ for $X\in\Map(\R^3;\g)$ and
$F_A-\eps$ for $A\in\cA(\R^3)$,
\alpheqn
\eqa
\label{LSa}
\Sym{\ccr{\eps}{X}}(\vp,\vx) &=& (-\ii) \f{\partial
\eps(\vp)}{\partial p_i}\partial_i X(\vx) + \OO(p^{-2}) \\
\label{LSb}
\Sym{F_A-\eps}(\vp,\vx) &=& \f{\partial
\eps(\vp)}{\partial p_i} A_i(\vx)  + \OO(p^{-2}).
\eqaend
\reseteqn
(Eq.\ \Ref{LSa} immediately follows from \Ref{as}.  An elementary argument
proving \Ref{LSb} is as follows.  One writes
$F_A=\D_A\slash\sqrt{\D_A^{\;2}}$.  Then with $\sig(\D_A)(\vp,\vx) =p\slD +
A\slD(\vx)$ one gets
$\sig(\D_A^{\;2})(\vp,\vx) = p^2\left(1 +\f{p\slD A\slD(\vx)}{p^2} +
\f{A\slD(\vx) p\slD}{p^2} + \OO(p^{-2})\right)$, and using \Ref{as},
\[
\sig(1/\sqrt{\D_A^{\;2}})(\vp,\vx) =
\f{1}{p}\left(1 -\half\f{p\slD A\slD(\vx)}{p^2} -
\half\f{A\slD p\slD(\vx)}{p^2}+ \OO(p^{-2})\right)
\]
implying
\[
\sig(F_A)(\vp,\vx) = \f{1}{p}\left(p\slD +A\slD(\vx)
 -\half\f{p\slD^2A\slD(\vx)}{p^2} -
\half\f{p\slD A\slD(\vx)p\slD}{p^2}\right) + \OO(p^{-2}).
\]
Noting that $\eps(\vp)=\f{p\slD}{p}$ and
$\f{\partial\eps(\vp)}{\partial p_i} = \f{1}{p}\left(\sigma_i -
\f{p\slD p_i}{p^2}\right)$, eq.\ \Ref{LSb} follows from the properties of the
Pauli matrices $\sig_i$.)

We will have to evaluate (regularized) traces only for operators with symbols
having compact support in $\vx$.  Note that such an operator $a$ is trace
class if and only
if its symbol is $\OO(p^{-4})$, and if it is not trace class we still can
define its {\em regularized trace} by introducing a momentum cut-off $\Lam>0$
for the divergent parts of its symbol,
\eqa
\label{TraL}
\TraL{a} =
\int_{p\leq\Lam}\dvp\intS\dvx\, \tra{\sig(a)(\vp,\vx)} +\nonu
\int_{p>\Lam}\dvp\intS\dvx\,
\tra{\left(\sig(a)-\sum_{j=0}^3\sig_{-j}(a)\right)(\vp,\vx)}.
\eqaend
Using the rules for symbol calculus above one can easily convince oneselves
that
\eq
\label{prop}
\TraL{a}=\TraL{\eps a\eps}
\eqend
for all bounded PDOs $a$ with symbols having compact support in $\vx$.
Obviously $\TraL{a}=\Tra{a}$ independent of $\Lam>0$ for trace class
operators $a$, and this implies
\eq
\label{TraC}
\TraC{a}= \TraL{a}\quad \mbox{$\forall \Lam>0$ if $a$ is conditionally trace
class}.
\eqend

\pgf{b}
For trace class operators $u,v$, the MR cocycle \cite{MR} is trivial and
can be represented as (see e.g.\ \cite{L}; this statement will be also
verified during our calculation below)
\eq
\label{cb}
c_{MR}(u,v;F) = (\delta b)(u,v;F)
\eqend
with
\eqa
b&=&b_1+b_2\nonu
b_1(u;F) &=& -\f{1}{2}\Tra{u\eps} \quad \mbox{ independent of $F$}\nonu
b_2(u;F) &=& \f{1}{16}\Tra{\ccr{\eps}{F}\ccr{\eps}{u}}.
\eqaend
The boundary operation $\delta$ is defined in (\ref{b1},b).
Defining $c_{MR}^\Lam$ as in \Ref{MR} with ${\rm Tr}_C$ replaced
by ${\rm Tr}_\Lam$ and similarly $b^\Lam$, $b_1^\Lam$ and $b_2^\Lam$,
we introduce
\eq
\Delta c_{MR}^\Lam  \equiv c_{MR}^\Lam - \delta b^\Lam .
\eqend
{}From \Ref{TraC} it follows that $c_{MR}(u,v;F)=c_{MR}^\Lam(u,v;F)$ for
$F\in\Grz$,
$u,v\in\gz$, hence we can write
\eq
c_{MR}(u,v;F) =  \Delta c_{MR}^\Lam (u,v;F) + (\delta b^\Lam)(u,v;F) .
\eqend

Eq.\ \Ref{cb} implies that for $u,v$ trace class we should get that $\Delta
c_{MR}^\Lam(u,v;F) = 0$.  We therefore expect that it should be possible to
make this explicit and represent $\Delta c_{MR}^\Lam (u,v;F)$ as a sum of terms
of the form $\TraL{\ccr{a}{b}}$ for some operators $a,b$ (note that
${\rm Tr}_\Lam$ is not cyclic: $\TraL{ab}\neq\TraL{ba}$ in general!).

Indeed, it is not difficult to find such a representation:
We write
\nonueqa
c_{MR}^\Lam&=& c^\Lam_1 + c^\Lam_2\nonu
c^\Lam_1(u,v;F) &=&
\f{1}{8}\TraL{\eps\ccr{\ccr{\eps}{u}}{\ccr{\eps}{v}}}\nonu
c^\Lam_2(u,v;F) &=&
-\f{1}{8}\TraL{F\ccr{\ccr{\eps}{u}}{\ccr{\eps}{v}}} .
\nonueqaend
We first calculate the $F$-independent part of $\Delta c_{MR}^\Lam$ and
obtain by a straightforward calculation using \Ref{prop}
\[
c^\Lam_1(u,v;F) - (\delta b^\Lam_1)(u,v;F) = \f{1}{4}\TraL{\ccr{u}{\eps v} -
(\cuv)}.
\]

To calculate the $F$ dependent part, we first observe that due to
\Ref{prop}, we can write $b_2^\Lam$ as
\[
b^\Lam_2(u;F) = -\f{1}{8}\TraL{F\eps\ccr{\eps}{u}}
\]
and that $(\delta b_2^\Lam)(u,v;F) = b_2^\Lam(\ccr{u}{v};F) -
b_2^\Lam(v;\ccr{F}{u})+b_2^\Lam(u;\ccr{F}{v})$ as $b_2^\Lam(u;F)$ is linear
in $F$; due to the Jacobi
identity, $\ccr{\eps}{\ccr{u}{v}}=\ccr{\ccr{\eps}{u}}{v} - (\cuv)$, hence
\[
(\delta b^\Lam_2)(u,v;F) = -\f{1}{8}\TraL{ F\eps\ccr{\ccr{\eps}{u}}{v} +
\ccr{F}{v}\eps\ccr{\eps}{u} - (\cuv)}
\]
Writing now
\[
c^\Lam_2(u,v;F) =
\f{1}{8}\TraL{F\ccr{\eps}{v}\ccr{\eps}{u} - (\cuv)}
\]
we see that, using the Jacobi identity for the commutator twice, we
can write
\[
c^\Lam_2(u,v;F) - (\delta b^\Lam_2)(u,v;F) =
\f{1}{8}\TraL{\ccr{F\eps\ccr{\eps}{u}}{v} - (\cuv)}.
\]
As $\TraL{\ccr{\ccr{\eps}{u}}{v} - (\cuv)} =
\TraL{\ccr{\eps}{\ccr{u}{v}}} = 0$ (we used the Jacobi identity and
\Ref{prop}), replacing $F$ in this expression by $(F-\eps)$
does not have any effect.

Collecting terms, we therefore obtain
\eqa
\label{ccr}
\Delta c_{MR}^\Lam(u,v;F) =
\f{1}{4}\TraL{\ccr{u}{\eps v} - (\cuv)}\nonu
+ \f{1}{8}\TraL{\ccr{(F -\eps) \eps\ccr{\eps}{u}}{v} - (\cuv)}.
\eqaend
which now is of the form we were after.

We now claim: For $X,Y\in\Map(\R^3;\g)$, $F_A=\sign(\D_A)$ with
$A\in\cA(\R^3)$, we have
\eqa
\label{equiv}
\Delta c^\Lam(X,Y;F_A) =
\f{\ii}{24\pi^2}\intS\dvx \, \trac{\epsilon_{ijk} A_i(\vx) (
\partial_j u(\vx) \partial_k v(\vx) - \partial_j v(\vx) \partial_k
u(\vx) )} \nonu = c_{MFS}(X,Y;A)
\eqaend
where $c_{MFS}$ is the MFS cocycle \Ref{MFS} ($\epsilon_{ijk}$ is the
antisymmetric tensor with $\epsilon_{123}=1$).

\pgf{c} To evaluate the l.h.s. of \Ref{equiv} we use symbol calculus.

As $\tras{\sigma_i}=0$, we obviously have $\TraL{\ccr{u}{\eps v}}=0$,
and the $F$-independent term in
\Ref{ccr} does not contribute to the l.h.s. of \Ref{equiv}.
Moreover,
\eq
\label{25}
\Sym{\ccr{(F_A-\eps)\eps\ccr{\eps}{X}}{Y}}(\vx,\vp)  =
(-\ii)^2 \f{\partial}{\partial p_i}\left(\f{\partial\eps(\vp)}{\partial
p_j}A_j(\vx)  \eps(\vp) \f{\partial\eps(\vp)}{\partial p_k}\partial_k
X(\vx)\right)\partial_i Y(\vx) +
\OO(p^{-4}) .
\eqend
Under ${\rm Tr}_\Lam$ the $\OO(p^{-4})$-term does not contribute,
hence we get
\eq
\label{ess1}
\Delta c^\Lam(X,Y;F_A) = \f{1}{8}J_{ijk}^\Lam
\intS\dvx\,\trac{A_j(\vx)\left(\partial_jX(\vx)\partial_kY(\vx) - (\cXY)
\right)}
\eqend
where
\eq
\label{J}
J_{ijk}^\Lam = -\int_{p\leq\Lam}\dvp
\tras{\f{\partial}{\partial p_i}\left( \f{\partial\eps(\vp)}{\partial
p_j}\eps(\vp) \f{\partial\eps(\vp)}{\partial p_k}\right)} .
\eqend
To evaluate the last integral, we note that
$\partial\eps(\vp)/\partial p_i = P_{il}\sig_l/p$ with $P_{il}=
(\delta_{il} - p_ip_l/p^2)$, hence with $\tras{\sig_i\sig_j\sig_k} =
2\ii\epsilon_{ijk}$ we get after a simple calculation
\[
J_{ijk}^\Lam = -\int_{p\leq\Lam}\dvp
\f{\partial}{\partial p_i} \left(2\ii \epsilon_{jlk}\f{p_l}{p^3}\right)
\]
and are left with the elementary integral
\[
I_{il}^\Lam = \int_{p\leq\Lam}\dvp \f{\partial}{\partial p_i}
\left(\f{p_l}{p^3}\right) = \f{\delta_{il}}{6\pi^2}.
\]
(The latter equality can be seen by an elementary calculation ($(ijk)$
is a cyclic permutation of $(1,2,3)$, and $(p_j,p_k) = (q\cos(\phi),
q\sin(\phi))$ polar coordinates as usual):
\nonueqa
I_{il}^\Lam = \f{1}{(2\pi)^3} \int_0^\Lam dq q \int_0^{2\pi}
d\phi \int_{-\sqrt{\Lam^2 - q^2}}^{\sqrt{\Lam^2 - q^2}}dp_i
\f{\partial}{\partial p_i}
\left(\f{p_l}{p^3}\right) \\ =
\f{1}{(2\pi)^3} \int_0^\Lam dq q \, 2 \pi
\, 2\delta_{il} \f{\sqrt{\Lam^2 -q^2}}{\Lam^3}
= \f{\delta_{il}}{6\pi^2} ;
\nonueqaend
note that $I_{ik}^\Lam$ does not depend on $\Lam$.) With that we get
\eq
J^\Lam_{ijk} = \f{\ii}{3\pi^2} \epsilon_{ijk}
\eqend
independent of $\Lam$ which together with eq.\ \Ref{ess1} proves the assertion.

The infrared singularity of the symbols (the pole at $p=0$) in the
calculation above is essential.  If one ignores it, the result is zero: If
one first performs the angular integration in momentum space in eq.\
\Ref{J} the trace ${\rm tr}_{spin}$ vanishes.
In the above calculation of $J_{ijk}^\Lam$ we have respected the
distributional nature of the momentum space derivatives of $\eps(\vp)$.
Another check for the computation is obtained if one replaces $\eps$ by a
smooth (non-singular) function $\tilde\eps(\vp)$ such that
$\tilde\eps(\vp)=\eps(\vp)$ far away from the origin.  The above
computation, when repeated for $\tilde\eps(\vp)$, shows that the trace is a
boundary integral in momentum space and {\em in no way depends on the
choice of the smoothing} $\tilde\eps(\vp)$ near the origin.  Thus the
commutator anomaly is a result of a nontrivial interplay of the ultraviolet
behavior and the infrared properties of gauge currents.

{\bf 3.  Final Remarks.} Given a Hilbert space $h$ and a grading operator
$\eps$ on $h$, one basic object of NCG is the graded differential complex
$\Om_p=\oplus_{n=0}^\infty\Om_p^{(n)}$ where $\Om_p^{(0)}=\gp$ and
$\Om_p^{(n)}$ is generated by linear combinations of operators of the form
\eq
\om_n=u\ccr{\eps}{v_1}\ldots\ccr{\eps}{v_n}\quad u,v_1,\ldots v_n\in\gp .
\eqend
Then
\eq
\ddd\om_n=\left\{\bma{lr} \ii\ccr{\eps}{\om_n}& \mbox{ if $n$ is even}\\
\ii\car{\eps}{\om_n}& \mbox{ if $n$ is odd} \ema\right.
\eqend
defines a derivation on $\Om$ satisfying $\ddd^2=0$ and which is supposed
to generalize the exterior derivative $\dd$ acting on $\g$-valued forms on a
$d$-dimensional manifold $M^d$, \cite{Connes}.

In noncommutative geometry one replaces the classical
$\gl(N)$-valued forms on $M^d$ in $\Om_p$ by operators by setting
\alpheqn
\eqa
\bma{ccl}\mbox{$\g$-valued form on $M^d$} &\to& \Om_p\\
X\dd Y_1\cdots \dd Y_n &\to& \ii^n
X\ccr{\eps}{Y_1}\cdots\ccr{\eps}{Y_n}\ema
\nonu \label{embeda}\forall X,Y_1,\ldots Y_n\in \Map(M^d;\gl(N))\subset\gp.
\eqaend
(Note that the arrow above is not really a well-defined map because a
vanishing linear combination of the classical forms on the left could lead
to a non-vanishing operator form on the right.) With this in mind, the
equations (\ref{LSa},b) for the symbols of operators look very suggestive:
it seems that the leading powers of operator symbols exactly realize the
embedding \Ref{embeda} for $p=2$ by assigning
\eq
\label{embedb}
\dd x^i\to \f{\partial\eps(\vp)}{\partial p_i}.
\eqend
Moreover it seems that $\cA(M^3)\ni A\to(F_A-\eps)$ is just a special case
of the embedding \Ref{embeda} for 1-forms.  The latter is not true, however:
if it was true, $\car{\eps}{F_A-\eps}+(F_A-\eps)^2=F_A^2-1$ should
correspond to the magnetic field $B\equiv (-\ii)\dd A+A^2$, but
$F_A^2-1=0$ always.  One can, however, find for every $A\in\cA(M^3)$ an
operator
$\Phi_A$ in $\Om_2^{(2)}$ whose symbol is identical to the one of $F_A$ to
leading- and next-to-leading order in $p^{-1}$, but which is such that
$\Phi_A^2-1$
is non-zero in general and  naturally represents $B$,
\[
\sig(\Phi_A^2-1)(\vp,\vx) = \f{\partial\eps(\vp)}{\partial p_i}
\f{\partial\eps(\vp)}{\partial p_j}\left(\partial_i A_j (\vx) -\partial_j
A_i (\vx)
+\ii\ccr{A_i(\vx)}{A_j(\vx)}\right) +\OO(p^{-3})
\]
(this was pointed out already by Rajeev \cite{Rajeev}).

The formula \Ref{MR} for the MR cocycle can therefore be regarded just
as the non-commutative generalization of the MFS cocycle \Ref{MFS} if one
replaces
\eq
\label{NCint}
\f{\ii}{3\pi^2} \int_{M^3} \to {\rm Tra}_C,
\eqend
\reseteqn
i.e.\ one regards the conditional trace as a non-commutative generalization
of integration of 3-forms on $M^3$. This is indeed very natural as one
can prove that \cite{LM}
\eqa
\f{\ii}{3\pi^2}\int_{M^3}\trac{X\dd Y_1\dd Y_2\dd
Y_3}=\ii^3\TraC{X\ccr{\eps}{Y_1}\ccr{\eps}{Y_2}\ccr{\eps}{Y_3
}}\nonu
\quad \forall X,Y_1,Y_2,Y_3\in\Map(M^3;\gl(N)) .
\eqaend
(There is an analog relation for arbitrary dimension $d$ \cite{LM}.  Note
that the resulting non-commutative integration calculus is different from
the one suggested in \cite{Connes1}.) This implies especially that the MFS
cocycle is {\em identical} to the MR cocycle in case $A$ is a pure gauge.
In general, one has an exact formula
\eq
c_{MFS}(X,Y,A)=c_{MR}(X,Y;\Phi_A)
\eqend
for a suitable choice of $\Phi_A$ as discussed above.


\bc{\bf Acknowledgments}\ec
We would like to thank S.G.  Rajeev for collaboration in an initial stage
of this work.  E.L. is grateful to G.W. Semenoff for helpful discussions
and would like to thank the Erwin Schr\"odinger International Institute in
Vienna for hospitality where part of this work was done.



\begin{thebibliography}{99}

\bibitem{LS} Langmann E. and Semenoff G. W.,
QCD(1+1) with massless quarks and gauge covariant Sugawara construction,
{\tt hep-th\slash 9404159}.
\bibitem{M} Mickelsson J., \LMP{7}, 45 (1983); \CMP{97}, 361 (1985).
\bibitem{FS} Faddeev L. D., \PL{145 B}, 81 (1984); Faddeev L. D. and
Shatashvili S. L., {\em Theor. Math. Phys.} {\bf 60}, 770 (1984).
\bibitem{MR} Mickelsson J. and Rajeev S. G., \CMP{116}, 400 (1988).
\bibitem{L} Langmann E., \CMP{162}, 1 (1994).
\bibitem{Connes} Connes A., {\em Publ. Math. IHES} {\bf 63}, 257 (1985).
\bibitem{RS1} Reed R., and Simon B.:
{\em Methods of Modern Mathematical Physics I. Functional
Analysis}, Academic Press, New York (1968); {\em II. Fourier Analysis,
Self-Adjointness}, Academic Press, New York (1975).
\bibitem{M3} Mickelsson J., \CMP{127}, 285 (1990).
\bibitem{symbol} H\"ormander L.: \em The Analysis of Linear Partial
Differential Operators III. \rm Springer-Verlag, Berlin (1985).
\bibitem{M1} Mickelsson J., Regularization of current algebra,
{\tt hep-th\slash 9306133}; Wodzicki residue and anomalies of
current algebras, {\tt hep-th\slash 9404093}; \LMP{28}, 97 (1993).
\bibitem{BJL} Jo S.-G., \PL{163B}, 353 (1985); Alekseev A., Madaichik Ya.,
Faddeev L.  D., and Shatashvili S. L., {\em Theor.  Math.  Phys.} {\bf 73},
1149 (1988).

\bibitem{Berry} Niemi A. J. and Semenoff G. W., \PRL{55}, 927 (1985);
Sonoda H., \NP{B266}, 410 (1986).
\bibitem{JJ} Jackiw R. and Johnson K., \PR{182}, 1459 (1969).
\bibitem{Rajeev} Rajeev S. G., \PR{D42}, 2779 (1990);
\PL{209B}, 53 (1988).
\bibitem{Rajeev1} Rajeev S. G., \PR{D44}, 1836 (1991).
\bibitem{LM} Langmann E. and Mickelsson J., in preparation.
\bibitem{Connes1} Connes A., \CMP{117}, 117 (1988).













\end{thebibliography}
\end{document}